\documentclass[preprint,showpacs,onecolumn,nofootinbib]{revtex4}
\pdfoutput=1
\usepackage[colorlinks=true,linkcolor=blue,urlcolor=blue,filecolor=black,citecolor=red,pdfstartview=FitV,pdftitle={},pdfsubject={},pdfkeywords={},pdfpagemode=None,bookmarksopen=true]{hyperref}
\usepackage{graphicx}
\usepackage{amsfonts}
\usepackage{amssymb,ulem}
\usepackage{color}%
\usepackage{dcolumn}
\usepackage{amsmath}
\usepackage{verbatim}
\usepackage{epsfig}
\usepackage{graphics}
\usepackage[active]{srcltx}
\usepackage{epstopdf}
\usepackage{shuffle}
\usepackage[utf8]{inputenc}

\newcommand{\bea}{\begin{eqnarray}}
\newcommand{\eea}{\end{eqnarray}}
\newcommand{\bean}{\begin{eqnarray*}}
\newcommand{\eean}{\end{eqnarray*}}
\newcommand{\nn}{\nonumber \\}

\def\Tr{\mathop{\rm Tr}}

\def\eref#1{(\ref{#1})}

\def\a{{\alpha}}

\def\b{{\beta}}

\def\Label#1{\label{#1}%
  \smash{\hbox to0pt{\raise1ex\hbox{\tiny[#1]}\hss}}}

\begin{document}

\baselineskip=0.6 cm
\title{
Recursive construction for expansions of tree Yang-Mills amplitudes from soft theorem}
\author{Chang Hu$^{ab}$, Kang Zhou$^{c}$}

\email{isiahalbert@126.com,zhoukang@yzu.edu.cn}

\affiliation{$^a$\small School of Fundamental Physics and Mathematical Sciences, Hangzhou Institute for Advanced Study, UCAS, Hangzhou 310024, China\\
$^b$ University of Chinese Academy of Sciences, Beijing 100049, China\\
$^c$ Center for Gravitation and Cosmology, College of Physical Science and Technology, Yangzhou University, Yangzhou, 225009, China}

\begin{abstract}
\baselineskip=0.6 cm

In this paper, we have introduced a fundamentally different approach, based on a bottom-up methodology, to expand tree-level Yang-Mills (YM) amplitudes into Yang-Mills-scalar (YMS) amplitudes and Bi-adjoint-scalar (BAS) amplitudes. Our method relies solely on the intrinsic soft behavior of external gluons, eliminating the need for external aids such as Feynman rules or CHY rules. The recursive procedure consistently preserves explicit gauge invariance at every step, ultimately resulting in a manifest gauge-invariant outcome when the initial expression is already framed in a gauge-invariant manner. The resulting expansion can be directly analogized to the expansions of gravitational (GR) amplitudes using the double copy structure. When combined with the expansions of Einstein-Yang-Mills amplitudes obtained using the covariant double copy method from existing literature, the expansions presented in this note yield gauge-invariant BCJ numerators.

\end{abstract}

\maketitle


\section{Introduction}
\label{sec-intro}

In recent decades, significant progress has been made in understanding scattering amplitudes, revealing hidden mathematical structures that go beyond traditional Feynman rules. These advancements have uncovered interconnections between amplitudes from various theories. For instance, gravitational (GR) and Yang-Mills (YM) amplitudes at the tree level are linked through the Kawai-Lewellen-Tye (KLT) relation \cite{Kawai:1985xq} and the Bern-Carrasco-Johansson (BCJ) color-kinematic duality \cite{Bern:2008qj,Chiodaroli:2014xia,Johansson:2015oia,Johansson:2019dnu}. In the well known CHY formalism \cite{Cachazo:2013gna,Cachazo:2013hca, Cachazo:2013iea, Cachazo:2014nsa,Cachazo:2014xea}, tree amplitudes for the a large variety of theories can be generated from the tree GR amplitudes through the compactifying, squeezing and the generalized dimensional reduction procedures\cite{Cachazo:2013gna,Cachazo:2013hca, Cachazo:2013iea, Cachazo:2014nsa,Cachazo:2014xea}. Similar unifying relations have emerged, allowing tree amplitudes from one theory to be expressed in terms of those from others \cite{Cheung:2017ems,Zhou:2018wvn,Bollmann:2018edb,Fu:2017uzt,Teng:2017tbo,Du:2017kpo,Du:2017gnh,Feng:2019tvb,Zhou:2019gtk,Zhou:2019mbe}.

The expansions of tree-level gravitational (GR) amplitudes into Yang-Mills (YM) amplitudes have garnered significant attention due to their role in constructing BCJ numerators \cite{Bern:2010yg,Fu:2017uzt}. These expansions involve recursive techniques, breaking down GR amplitudes into linear combinations of tree-level Einstein-Yang-Mills (EYM) amplitudes where some gravitons become gluons. They also expand tree-level EYM amplitudes into versions with fewer gravitons and more gluons \cite{Fu:2017uzt,Teng:2017tbo,Feng:2019tvb}. By iteratively applying these recursions, one can derive the desired GR to YM expansions.

The obtained coefficients or BCJ numerators through the aforementioned process are local, without any spurious poles in the expansions. However, establishing gauge invariance is challenging. Recent developments in scattering amplitude research have shown that amplitudes manifesting gauge invariance at the expense of locality often lead to innovative insights and unexpected mathematical structures. For instance, the Britto-Cachazo-Feng-Witten (BCFW) on-shell recursion relations\cite{Britto:2004ap,Britto:2005fq}, inspiring novel amplitude constructions like the Grassmannian representation and the Amplituhedron \cite{Arkani-Hamed:2012zlh,Arkani-Hamed:2013jha,Arkani-Hamed:2013kca}. Thus, exploring expansion techniques that explicitly showcase gauge invariance is valuable.

To achieve these new expansions, a natural approach is to modify existing recursive expansions. Fortunately, we have alternative recursive expansions for tree Einstein-Yang-Mills (EYM) amplitudes that explicitly exhibit gauge invariance for each polarization carried by external particles. These new expansions, discovered by Clifford Cheung and James Mangan using the covariant color-kinematic duality \cite{Cheung:2021zvb}, can also be derived using a technique based on soft theorems, similar to the one presented in this paper \cite{Wei:2023yfy}. Another challenge is to discover new expansions that manifest gauge invariance while expanding tree GR amplitudes into EYM ones. This can be achieved by altering the old GR expansions through the addition of new terms that vanish due to the gauge invariance of the expanded GR amplitudes, as outlined in subsection IV A. Ensuring that the original expansions describe gauge-invariant amplitudes is essential, making it logical to derive the old expansions from a gauge-invariant foundational framework, such as the traditional Lagrangian or CHY formalism. In other words, modifying the old expansions is a top-down construction.

The central aim of the modern S-matrix program is to construct amplitudes from a bottom-up approach, free from reliance on Lagrangian techniques. One prominent example involves bootstrapping 3-point amplitudes within the spinor-helicity formalism and subsequently generating higher-point amplitudes through the application of the BCFW recursive method \cite{Britto:2004ap,Britto:2005fq}. The primary objective of this paper is to provide a concise bottom-up approach for constructing expansions of tree-level gravitational (GR) amplitudes into tree-level Einstein-Yang-Mills (EYM) amplitudes. This approach ensures the explicit manifestation of gauge invariance for the polarization of each external particle.

The method presented in this paper is based on the sub-leading soft behavior for external gluons. Initially, soft theorems at the tree level were derived using Feynman rules for photons and gravitons \cite{Low,Weinberg}. In 2014, new soft theorems were discovered for gravity (GR) and Yang-Mills (YM) theory at the tree level by applying BCFW recursion relations \cite{Cachazo:2014fwa,Casali:2014xpa}. In GR, the soft theorem was extended beyond the leading order to sub-leading and sub-sub-leading levels, while in YM theory, it was identified at the leading and sub-leading orders. These new soft theorems were subsequently generalized to arbitrary spacetime dimensions \cite{Schwab:2014xua,Afkhami-Jeddi:2014fia} using CHY formulas \cite{Cachazo:2013gna,Cachazo:2013hca, Cachazo:2013iea, Cachazo:2014nsa,Cachazo:2014xea}. These soft theorems have been instrumental in constructing tree amplitudes, such as through the inverse soft theorem program, and by utilizing another type of soft behavior known as the Adler zero to construct amplitudes for various effective theories \cite{Cheung:2014dqa,Luo:2015tat,Elvang:2018dco,Cachazo:2016njl,Rodina:2018pcb,Boucher-Veronneau:2011rwd,Nguyen:2009jk}. In this paper, we build upon the idea of constructing tree amplitudes from soft theorems, but our approach differs significantly from the techniques found in existing literature.

Instead of expanding GR (gravity) amplitudes to EYM (Einstein-Yang-Mills) ones, this paper focuses on the expansions of color-ordered single-trace tree YM (Yang-Mills) amplitudes \cite{Chiodaroli:2014xia,Chiodaroli:2017ngp} into tree Yang-Mills-scalar (YMS) amplitudes,  which describe gluons and bi-adjoint scalars (BAS). This choice is motivated by several factors. Firstly, both expansions share the same coefficients due to the double copy structure \cite{Kawai:1985xq,Bern:2008qj,Chiodaroli:2014xia,Johansson:2015oia,Johansson:2019dnu,Cachazo:2013iea}. Secondly, the pure BAS amplitudes only contain propagators without numerators \cite{Cachazo:2013iea}. The approach starts by bootstrapping the lowest 3-point tree YM amplitudes, while assuming that higher-point tree amplitudes are uniquely determined by the soft behaviors of gluons. The advantage lies in the sub-leading soft factor of gluons, which inserts the soft external leg in a manifestly gauge-invariant manner. Consequently, if the original 3-point amplitude is expressed in a manifestly gauge-invariant formula, this gauge invariance is preserved throughout the process. As demonstrated, relying on the assumption that soft behaviors fully determine amplitudes, gauge invariance is established for the old expansions without necessitating a top-down derivation.

The method presented in this paper represents a significant advancement and improvement compared to the one used in \cite{Zhou:2022orv}. The new version constructs general expansions starting from the lowest-point amplitudes, which can be determined through bootstrapping. As pointed out in \cite{Zhou:2022orv}, the soft theorem for gluons can be uniquely fixed by assuming the universality of soft behaviors. More explicitly, one can find the expansion of tree YMS amplitudes with one external gluon by imposing the universality of the soft behaviours of BAS scalars, then derive the soft theorem for gluons from the resulted expansion of such YMS amplitudes. On the other hand, since the pure BAS amplitudes only contain propagators, the soft behaviors of BAS scalars are obvious. Thus, in the whole story, the expansions of YM amplitudes to YMS ones indeed arise from the assumptions of universality of soft behaviors.

The remainder of this note is organized as follows. In section \ref{sec-background}, we introduce necessary background including the expansions of tree amplitudes to KK BAS basis, the recursive expansions of tree YMS (EYM) amplitudes, as well as the soft theorems for external BAS scalars and gluons at tree level. In section \ref{sec-recur-constru}, we introduce our recursive method, and construct the old expansions of tree YM amplitudes to YMS ones, from the $3$-point amplitudes fixed by bootstrapping. In section \ref{sec-GI}, we derive the new expansions which manifest the gauge invariance, by applying both the direct construction and the recursive technic. Finally we end with a brief summery in section \ref{sec-conclusion}.

\section{Background}
\label{sec-background}

For readers' convenience, in this section we give a brief review of necessary background. In subsection \ref{subsecexpand}, we introduce the concept of tree level amplitudes of bi-adjoint scalar (BAS) theory and Yang-Mills scalar (YMS) theory. Meanwhile, we draw the conclusion that any theory consisting solely of massless particles can be expanded onto the KK BAS basis.  In subsection \ref{subsec-soft}, We briefly list the previously established soft theorems for scalars and gluons in various theories. In subsection \ref{subsec-expan-YMS}, We provide a concise introduction on how to recursively expand YMS tree-level amplitudes into the KK basis.

\subsection{Expanding tree level amplitudes to BAS basis}
\label{subsecexpand}

\begin{figure}
  \centering

   \includegraphics[width=15cm]{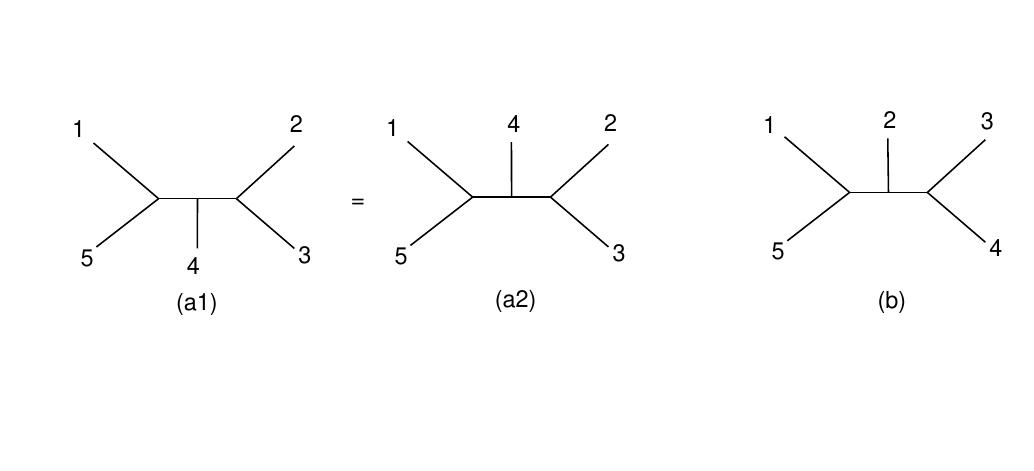}  \\
  \caption{Two $5$-point diagrams. Figure (a1) and (a2) are the same tree diagram, while (a1) represents the color ordering $12345$ and (a2) represents the color ordering $14235$. }\label{5p}
\end{figure}

The double color-ordered tree amplitudes in bi-adjoint scalar (BAS) theory exclusively feature propagators for massless scalars. Each amplitude exhibits simultaneous planarity with respect to two color orderings. We take the $5$-point amplitude ${\cal A}_{\rm BAS}(1,2,3,4,5|1,4,2,3,5)$ as an example. In Figure \ref{5p}, both $(a1)$ and $(a2)$ correspond to the same tree diagram. Specifically, a1 is associated with color ordering $(1,2,3,4,5)$, while $(a2)$ corresponds to color ordering $(1,4,2,3,5)$. However, in Figure \ref{5p}, the tree diagram represented by figure $b$ can only conform to color ordering $(1,2,3,4,5)$ and does not satisfy color ordering $(1,4,2,3,5)$. Similarly, one can draw additional tree diagrams that satisfy color ordering $(1,2,3,4,5)$, but it will be observed that none of them conform to color ordering $(1,4,2,3,5)$. Then the tree BAS amplitude ${\cal A}_{\rm BAS}(1,2,3,4,5|1,4,2,3,5)$ can be computed as
\bea
{\cal A}_{\rm BAS}(1,2,3,4,5|1,4,2,3,5)=(-1)^{n_{flip}}{1\over s_{23}}{1\over s_{51}}\,.
\eea
The Mandelstam variable $s_{i\cdots j}$ is defined as%
\bea
s_{i\cdots j}\equiv k_{i\cdots j}^2\,,~~~~k_{i\cdots j}\equiv\sum_{a=i}^j\,k_a\,,~~~~\label{mandelstam}
\eea
where $k_a$ is the momentum carried by the external leg $a$.

Each double color-ordered BAS amplitude carries an overall sign $(-1)^{n_{flip}}$, where $n_{flip}$ is determined by these two color orderings. Readers should be aware that in this paper, when these two color orderings are identical, $n_{flip}$ equals zero, implying that the overall sign is $+$. The systematic method for determining $n_{flip}$ and evaluating double color ordered partial amplitudes can be referenced in \cite{Cachazo:2013iea}, and as it is not pertinent to the current paper, it will not be further elaborated upon here. \footnote{In fact, in reference A, the overall sign is defined as $(-1)^{n-3+n_{flip}}$, where $n$ is the number or external legs. However, for the convenience of our subsequent work in this paper, we adopt the notation $(-1)^{n_{flip}}$. The advantage of the new convention stems from the fact that when we remove a soft external scalar from an $n$-point amplitude, it changes the number of external legs to $n-1$. Therefore, we desire that this overall sign contains information solely about the relative ordering between the two color orderings of the origin $n$-point amplitude, excluding the influence of changes in the number of external legs on this sign.}

The Yang-Mills-scalar (YMS) amplitudes pertain to scalars that are coupled with gluons. In this context, we are specifically concerned with a subset of YMS amplitudes known as single-trace YMS amplitudes. The single-trace YMS amplitude, denoted as ${\cal A}_{\rm YMS}(1,\cdots,n;\{p_1,\cdots,p_m\}|\sigma_{n+m})$, comprises of $n$ external scalars represented by $\{1,\cdots,n\}$ and $m$ external gluons labeled as $\{p_1,\cdots,p_m\}$. The ordering $\sigma_{n+m}$ on the right-hand side of $|$ encompasses all external legs within $\{1,\cdots,n\}\cup\{p_1,\cdots,p_m\}$. On the left-hand side of $|$, $(1,\cdots,n)$ signifies an alternative overall ordering of external scalars, which is also the reason it is referred to as 'single-trace', while $\{p_1,\cdots,p_m\}$ constitutes an unordered set. In simpler terms, the external scalars are BAS scalars, which have a dual ordering, whereas the external gluons belong to a single ordering, $\sigma_{n+m}$.

\begin{figure}
  \centering
   \includegraphics[width=13cm]{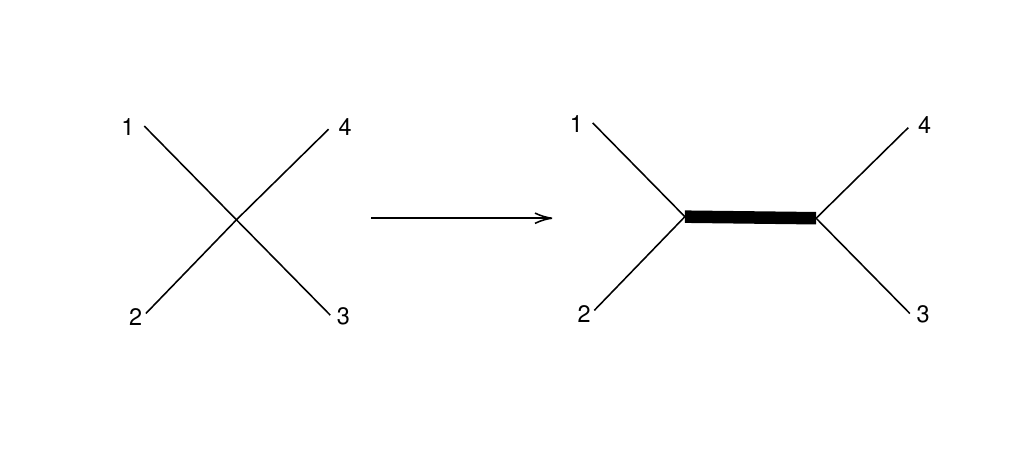}\\
  \caption{Turn the $4$-point vertex to $3$-point ones. The bold line corresponds to the inserted propagator $1/s_{12}$. This manipulation turns the original numerator $N$ to $s_{12}N$, and split the original coupling constant $g$ to two $\sqrt{g}$ for two cubic vertices.}
\label{3pto4p}
\end{figure}

Tree level amplitudes for any theory as long as they contain only massless particles and cubic vertices can be expanded to double color ordered BAS amplitudes,
due to the observation that each Feynman diagram for pure propagators is included in at least one BAS amplitude. In addition, for tree diagrams involving higher point vertices, we can decompose them into BAS amplitudes with solely cubic vertices by multiplying both the numerator and denominator by the propagator D.  An example is shown in Figure \ref{3pto4p}. Since each Feynman diagram contributes propagators that can be mapped to BAS amplitudes, along with a numerator, we can conclude that each tree-level amplitude can be expanded into double color-ordered BAS amplitudes. These expanded amplitudes have coefficients that are polynomials depending on Lorentz invariants created by external kinematic variables.

In reality, not all BAS amplitudes are independent, therefore, such an expansion requires the selection of appropriate basis.
Such basis can be determined by the well known Kleiss-Kuijf (KK) relation \cite{Kleiss:1988ne}
\bea
{\cal A}_{\rm BAS}(1,\vec{\pmb{\a}},n,\vec{\pmb{\b}}|\sigma_n)=(-)^{\vec{|\pmb{\b}}|}\,{\cal A}_{\rm BAS}(1,\vec{\pmb{\a}}\shuffle\vec{\pmb{\b}}^T,n|\sigma_n)\,,~~~\label{KK}
\eea
where $\vec{\pmb{\a}}$ and $\vec{\pmb{\b}}$ are two ordered subsets of external scalars, and $\vec{\pmb{\b}}^T$ stands for the ordered set generated from $\vec{\pmb{\b}}$ by reversing the original ordering, the $|\vec{\pmb{\b}}|$ on overall sign represents the number of elements in subset $\vec{\pmb{\b}}$. The symbol $\shuffle$ means summing over all
possible shuffles of two ordered sets $\vec{\pmb{\b}}_1$ and $\vec{\pmb{\b}}_2$, i.e., all permutations in the set $\vec{\pmb{\b}}_1\cup \vec{\pmb{\b}}_2$ while preserving the orderings
of $\vec{\pmb{\b}}_1$ and $\vec{\pmb{\b}}_2$. For instance, suppose $\vec{\pmb{\b}}_1=\{1,2\}$ and $\vec{\pmb{\b}}_2=\{3,4\}$, then
\bea
\vec{\pmb{\b}}_1\shuffle \vec{\pmb{\b}}_2&=&(1,2,3,4)+(1,3,2,4)+(1,3,4,2)+(3,1,2,4)+(3,1,4,2)+(3,4,1,2)\,.~~~~\label{shuffle}
\eea

One should note that
the $n$-point BAS amplitude ${\cal A}_{\rm BAS}(1,\vec{\pmb{\a}},n,\vec{\pmb{\b}}|\sigma_n)$ at the l.h.s of \eref{KK} carries two color orderings.
We can also have an another analogous KK relation for the second color ordering $\sigma_n$. Therefore,
the KK relation implies that the basis can be chosen as BAS amplitudes
${\cal A}_{\rm BAS}(1,\sigma_1,n|1,\sigma_2,n)$, with $1$ and $n$ are fixed at two ends in each ordering. Such basis is called the KK BAS basis. Based on the discussion above, any amplitude which includes only massless particles can be expanded to this basis\footnote{The well known Bern-Carrasco-Johansson (BCJ) relation \cite{Bern:2008qj,Chiodaroli:2014xia,Johansson:2015oia,Johansson:2019dnu} links BAS amplitudes in the KK basis together, and the independent BAS amplitudes can be obtained by fixing three legs at three particular positions in the color orderings. However, in the BCJ relation, coefficients of BAS amplitudes depend on Mandelstam variables, this character leads to poles in coefficients when expanding to BCJ basis. On the other hand, when expanding to KK basis, one can find the expanded formula in which the coefficients contain no poles. In this paper, we opt for the KK basis, as we aim to ensure that all poles of tree amplitudes are encompassed within the basis, with coefficients solely serving as numerators.}. In the expansion,
the KK basis supplies the propagators, while the coefficients in the expansions provide the numerators.

In this paper, we will concentrate on the expansion of pure YM amplitudes. The $n$-point YM amplitude ${\cal A}_{\rm YM}(\sigma_n)$ with color ordering $\sigma_n$ can be expanded to KK BAS basis as
\bea
{\cal A}_{\rm YM}(\sigma_n)
&=&\sum_{\alpha_{n-2}}\,{\cal C}(\alpha_{n-2},\epsilon_i,k_i)\,{\cal A}_{\rm BAS}(1,\alpha_{n-2},n|\sigma_n)\,,~~~\label{exp-YM-KK}
\eea
where $\alpha_{n-2}$ denotes color orderings among external legs of $\{2,\cdots,n-1\}$. The double copy structure \cite{Kawai:1985xq,Bern:2008qj,Chiodaroli:2014xia,Johansson:2015oia,Johansson:2019dnu} indicates that the coefficient ${\cal C}(\alpha_{n-2},\epsilon_i,k_i)$ depends on polarization vectors $\epsilon_i$ and
momenta $k_i$ of external gluons, as well as the ordering $\alpha_{n-2}$, but remains independent of ordering $\sigma_n$\footnote{Originally, the concept of the double copy implies that the General Relativity (GR) amplitude can be factorized as ${\cal A}{\rm G}={\cal A}{\rm YM}\times {\cal S}\times{\cal A}_{\rm YM}$, where the kernel ${\cal S}$ is derived by inverting BAS amplitudes. Our assumption that the coefficients depend on only one color ordering is consistent with the original version, as discussed in \cite{Zhou:2022orv}.}. Therefore, within this expansion, the investigation of the structure of YM amplitudes is transformed into the study of the structure of the coefficients ${\cal C}(\alpha_{n-2},\epsilon_i,k_i)$. These coefficients are known as BCJ numerators \cite{Bern:2010yg,Fu:2017uzt}.


\subsection{Soft theorems for external scalars and gluons}
\label{subsec-soft}

In this subsection, we provide a brief overview of the soft scattering theorems for external scalars and gluons, which are crucial for the discussions in the subsequent sections.

For the double color ordered BAS amplitude ${\cal A}_{\rm BAS}(1,\cdots,n|\sigma_n)$, we rescale $k_i$
as $k_i\to\tau k_i$, and expand the amplitude with respect to $\tau$,
\begin{equation}
    {\cal A}_{\rm BAS}(1,\cdots,n|\sigma_n)={\cal A}^{(0)_i}_{\rm BAS}(1,\cdots,n|\sigma_n)+{\cal A}^{(1)_i}_{\rm BAS}(1,\cdots,n|\sigma_n)+\mathcal{O}(\tau).
\end{equation}
The leading-order contribution ${\cal A}^{(0)_i}_{\rm BAS}(1,\cdots,n|\sigma_n)$ arises explicitly from the $2$-point channels $1/s_{1(i+1)}$ and $1/s_{(i-1)i}$, which are at the $1/\tau$ order. In other words,
\bea
{\cal A}^{(0)_i}_{\rm BAS}(1,\cdots,n|\sigma_n)&=&{1\over \tau}\Big({\delta_{i(i+1)}\over s_{i(i+1)}}+{\delta_{(i-1)i}\over s_{(i-1)i}}\Big)\,
{\cal A}_{\rm BAS}(1,\cdots,i-1,\not{i},i+1,\cdots,n|\sigma_n\setminus i)\nn
&=&S^{(0)_i}_s\,{\cal A}_{\rm BAS}(1,\cdots,i-1,\not{i},i+1,\cdots,n|\sigma_n\setminus i)\,,~~~\label{soft-theo-s}
\eea
where $\not{i}$ stands for removing the leg $i$, $\sigma_n\setminus i$ means the color ordering generated from $\sigma_n$ by eliminating $i$. In Equation \eqref{soft-theo-s}, $\delta_{ab}$ is not the well-known Kronecker symbol. Its value is determined by the second color ordering $\sigma_{n}$. When $a,b$ in $\sigma_n$ are not adjacent, $\delta_{ab}$ takes the value of $0$. When $a,b$ in $\sigma_n$ are adjacent and maintain the same order as in the first color ordering $(1,\cdots, n)$, $\delta_{ab}$ takes the value of $1$. When $a,b$ in $\sigma_n$ are adjacent but have the reverse order compared to their order in the first color ordering $(1,\cdots,n)$, it takes the value of $-1$. The leading soft operator $S^{(0)}_s(i)$ for the scalar $i$ is observed as
\bea
S^{(0)_i}_s\equiv{1\over \tau}\,\Big({\delta_{i(i+1)}\over s_{i(i+1)}}+{\delta_{(i-1)i}\over s_{(i-1)i}}\Big)\,.~~~~\label{soft-fac-s-0}
\eea

For example, for $4$-point amplitude ${\cal A}_{BAS}(1234|1234)$, we rescale $k_4$ as $k_4\to \tau k_4$, then by our notation, we have
\bea
        &{\cal A}_{BAS}^{(0)}(1234|1234)=&\frac{1}{\tau}(\frac{1}{s_{41}}+\frac{1}{s_{34}})\nn
        &{\cal A}_{BAS}(123|123)=&+1.
\eea
According to our definition of $\delta_{ab}$, we know $\delta_{34}=\delta_{41}=1$ since $3,4,1$ remains the same in both color orderings. So we verified

\bea
   {\cal A}^{(0)_4}_{\rm BAS}(1234|1234)&=&{1\over \tau}\Big({\delta_{41}\over s_{41}}+{\delta_{34}\over s_{34}}\Big)\,
{\cal A}_{\rm BAS}(123|123).
\eea
For $4$-point amplitude ${\cal A}_{BAS}(1234|1243)$, we also rescale $k_4$ as $k_4\to \tau k_4$, then by our notation, we have

\bea
        &{\cal A}_{BAS}^{(0)}(1234|1243)=&\frac{1}{\tau}(-\frac{1}{s_{34}})\nn
    &{\cal A}_{BAS}(123|123)=&+1.
\eea
Because 3 and 4 have opposite orders in the two color orderings $1234$ and $1243$, we have $\delta_{34}=-1$. And since 4 and 1 are not adjacent in the second color ordering $1243$, we have $\delta_{41}=0$. So we verified

\bea
   {\cal A}^{(0)_4}_{\rm BAS}(1234|1243)&=&{1\over \tau}\Big({\delta_{41}\over s_{41}}+{\delta_{34}\over s_{34}}\Big)\,
{\cal A}_{\rm BAS}(123|123).
\eea

In our previous work, we introduced an assumption that, the soft operator for should be universal across different theories. For example, in the YMS theory, scalar particles, and in the BAS theory, scalar particles, both share the same form for their respective soft operators. Then the leading contribution of YMS amplitude is
\bea
{\cal A}^{(0)_i}_{\rm YMS}(1,\cdots,n;p_1,\cdots,p_m|\sigma_{n+m})
&=&S^{(0)_i}_s\,{\cal A}_{\rm YMS}(1,\cdots,i-1,\not{i},i+1,\cdots,n;p_1,\cdots,p_m|\sigma_{n+m}\setminus i)\,.~~~\label{soft-theo-s2}
\eea
Based on \eqref{soft-fac-s-0}, it can be observed that this soft operator $S^{(0)_i}_s$ does not act on external gluons.

The soft theorems for external gluons at leading and sub-leading orders can be obtained via various approaches \cite{Casali:2014xpa,Schwab:2014xua,Afkhami-Jeddi:2014fia,Zhou:2022orv}. Such soft theorems are given as
\bea
{\cal A}^{(0)_{p_i}}_{\rm YMS}(1,\cdots,n;p_1,\cdots,p_m|\sigma_{n+m})
&=&S^{(0)_{p_i}}_g\,{\cal A}_{\rm YMS}(1,\cdots,n;p_1,\cdots,\not{p_i},\cdots,p_m|\sigma_{n+m}\setminus p_i)\,,~~~\label{soft-theo-g0}
\eea
and
\bea
{\cal A}^{(1)_{p_i}}_{\rm YMS}(1,\cdots,n;p_1,\cdots,p_m|\sigma_{n+m})
&=&S^{(1)_{p_i}}_g\,{\cal A}_{\rm YMS}(1,\cdots,n;p_1,\cdots,\not{p_i},\cdots,p_m|\sigma_{n+m}\setminus p_i)\,,~~~\label{soft-theo-g1}
\eea
where the external momentum $k_{p_i}$ is rescaled as $k_{p_i}\to\tau k_{p_i}$. The soft factors at leading and sub-leading orders are given by
\bea
S^{(0)_{p_i}}_g={1\over\tau}\,\sum_{a\in\sigma}\,{\delta_{ap_i}\,(\epsilon_{p_i}\cdot k_a)\over s_{ap_i}}\,,~~\label{soft-fac-g-0}
\eea
and
\bea
S^{(1)_{p_i}}_g=\sum_{a\in\sigma}\,{\delta_{ap_i}\,\big(\epsilon_{p_i}\cdot J_a\cdot k_{p_i}\big)\over s_{ap_i}}\,,~~\label{soft-fac-g-1}
\eea
respectively \cite{Casali:2014xpa,Schwab:2014xua}. Here, $\delta_{ab}$ is solely determined by the second color ordering $\sigma_{n+m}$. When $a,b$ are adjacent in $\sigma_{n+m}$ (in a cyclically symmetric sense), $\delta_{ab}$ takes the value of $1$ if $a$ precedes $b$, and $-1$ if $a$ follows $b$. When $a,b$ are not adjacent in $\sigma_{n+m}$, $\delta_{ab}$ takes the value of $0$.  $J_a^{\mu\nu}$ is the angular momentum operator, defined as $J_a^{\mu\nu}=k_a^{[\mu}\,{\partial \over\partial k_{a,\nu]}}$. In \eref{soft-fac-g-0} and \eref{soft-fac-g-1}, one should sum over all external legs $a$, i.e., these soft operators for external gluon act on both external scalars and gluons.

The sub-leading soft operator \eref{soft-fac-g-1} for external gluons plays a central role in the subsequent sections. Here, we present some valuable results regarding the operation of this operator. The angular momentum operator $J_a^{\mu\nu}$ acts on the Lorentz vector $k^\rho_a$ with the orbital component of the generator and on $\epsilon^\rho_a$ with the spin component of the generator in the vector representation.
\bea
J_a^{\mu\nu}\,k_a^\rho=k_a^{[\mu}\,{\partial k_a^\rho\over\partial k_{a,\nu]}}\,,~~~~
J_a^{\mu\nu}\,\epsilon_a^\rho=\big(\eta^{\nu\rho}\,\delta^\mu_\sigma-\eta^{\mu\rho}\,\delta^\nu_\sigma\big)\,\epsilon^\sigma_a\,.
\eea
Then the action of $S^{(1)_p}_g$ can be re-expressed as
\bea
S^{(1)_p}_g&=&-\sum_{V_a}\,{\delta_{ap}\over s_{ap}}\,V_a\cdot f_p\cdot{\partial\over\partial V_a}\,,~~\label{act-s-V}
\eea
due to the observation that the amplitude is linear in each polarization vector. We observe that in \eref{act-s-V}, the summation over $V_a$ includes all Lorentz vectors, encompassing both momenta and polarizations. \textbf{The operator \eref{act-s-V} is a differential operator that adheres to Leibniz's rule}. Employing \eref{act-s-V}, we promptly obtain
\bea
\big(S^{(1)_p}_g\,k_a\big)\cdot V=-{\delta_{ap}\over s_{ap}}\,(k_a\cdot f_p\cdot V)\,,~~~~\big(S^{(1)_p}_g\,\epsilon_a\big)\cdot V=-{\delta_{ap}\over s_{ap}}\,(\epsilon_a\cdot f_p\cdot V)\,,~~~~\label{iden-1}
\eea
where $V$ is an arbitrary Lorentz vector, and
\bea
V_1\cdot\big(S^{(1)_p}_g\,f_a\big)\cdot V_2={\delta_{ap}\over s_{ap}}\,V_1\cdot(f_p\cdot f_a-f_a\cdot f_p)\cdot V_2\,,~~~~\label{iden-2}
\eea
for two arbitrary Lorentz vectors $V_1$ and $V_2$, where the anti-symmetric tensor $f_i$ is defined as $f_i^{\mu\nu}\equiv k_i^\mu\epsilon_i^\nu-\epsilon_i^\mu k_i^\nu$, as introduced previously.

\subsection{Recursive expansion of single-trace YMS amplitudes}
\label{subsec-expan-YMS}

The discussion for the expansion of tree level amplitudes in the previous subsection.\ref{subsecexpand} indicates that the YMS amplitudes can also be expanded to KK BAS basis. This expansion can be achieved by applying the following recursive expansion iteratively:
\bea
& &{\cal A}_{\rm YMS}(1,\cdots,n;\{p_1,\cdots,p_m\}|\sigma_{n+m})\nn
&=&\sum_{\vec{\pmb{\a}}}\,\big(\epsilon_p\cdot F_{\vec{\pmb{\a}}}\cdot Y_{\vec{\pmb{\a}}}\big)\,{\cal A}_{\rm YMS}(1,\{2,\cdots,n-1\}\shuffle \{\vec{\pmb{\a}},p\},n;\{p_1,\cdots,p_m\}\setminus\{p\cup\pmb{\a}\}|\sigma_{n+m})\,,~~\label{expan-YMS-recur}
\eea
where $p$ is the fiducial gluon which can be chosen as any element in $\{p_1,\cdots,p_m\}$, and $\pmb{\a}$ are subsets of $\{p_1,\cdots,p_m\}\setminus p$ which is allowed to be empty. When $\pmb{\a}=\{p_1,\cdots,p_m\}\setminus p$, the YMS amplitudes in the second line of \eref{expan-YMS-recur} are reduced to BAS ones. The ordered set $\vec{\pmb{\a}}$ is generated from $\pmb{\a}$
by endowing an order among elements in $\pmb{\a}$. The tensor $F^{\mu\nu}_{\vec{\pmb{\a}}}$ is defined as
\bea
F^{\mu\nu}_{\vec{\pmb{\a}}}\equiv\big(f_{\a_k}\cdot f_{\a_{k-1}}\cdots f_{\a_2}\cdot f_{\a_1}\big)^{\mu\nu}\,,
\eea
for $\vec{\pmb{\a}}=\{\a_1,\cdots\a_k\}$. The combinatory momentum $Y_{\vec{\pmb{\a}}}$ is the summation of momenta carried by external gluons at the l.h.s of $\a_1$ in the color ordering $\{2,\cdots,n-1\}\shuffle \vec{\pmb{\a}}$. The summation in \eref{expan-form1} is over all un-equivalent ordered sets $\vec{\pmb{\a}}$.
The recursive expansion \eref{expan-YMS-recur} is found via various methods \cite{Casali:2014xpa,Schwab:2014xua,Afkhami-Jeddi:2014fia,Zhou:2022orv}. It should be noted that this recurrence relation can be independently derived through the recursive construction based on soft behaviors. For detailed information, please refer to the reference \cite{Zhou:2022orv}. In the recursive expansion \eref{expan-YMS-recur}, the YMS amplitude undergoes expansion, resulting in YMS amplitudes with fewer gluons and more scalars. By iteratively applying this expansion, it becomes possible to expand any YMS amplitude into pure BAS amplitudes.

Furthermore, we can also observe that in the recursive expansion \eref{expan-YMS-recur}, the gauge invariance for each gluon in $\{P_1,\cdots,p_m\}\setminus p$ is manifest, since the tensor $f^{\mu\nu}$ vanishes automatically under the replacement $\epsilon_i\to k_i$, due to the definition. However, the gauge invariance for the fiducial gluon $p$ has not been manifested. When applying \eref{expan-YMS-recur} iteratively, fiducial gluons will be chosen at every step. Consequently, in the resulting expansion to pure BAS amplitudes, the gauge invariance for each gluon will be compromised. To achieve a manifestly gauge-invariant expansion, an alternative recursive expansion should be employed

\bea
& &{\cal A}_{\rm YMS}(1,\cdots,n;\{p_1,\cdots,p_m\}|\sigma_{n+m})\nn
&=&\sum_{\vec{\pmb{\a}}}\,{k_r\cdot F_{\vec{\pmb{\a}}}\cdot Y_{\vec{\pmb{\a}}}\over k_r\cdot k_{p_1\cdots p_m}}\,{\cal A}_{\rm YMS}(1,\{2,\cdots,n-1\}\shuffle \vec{\pmb{\a}},n;\{p_1,\cdots,p_m\}\setminus\pmb{\a}|\sigma_{n+m})\,,~~\label{expan-YMS-recur-GI}
\eea
where $k_r$ is a reference massless momentum. In this context, the notations align with those in \eref{expan-YMS-recur}, and $\pmb{\a}$ represents subsets of ${p_1, \cdots, p_m}$. The formula \eref{expan-YMS-recur-GI} was originally discovered by Clifford Cheung and James Mangan within the framework of the covariant double copy. It was also independently derived by applying the recursive construction based on soft theorems[]. Notably, the expansion \eref{expan-YMS-recur-GI} doesn't necessitate the use of a fiducial gluon, and it inherently exhibits gauge invariance for all polarizations. Through iterative application of \eref{expan-YMS-recur}, one eventually arrives at the manifestly gauge-invariant expansion of YMS amplitudes into the KK BAS basis.

\section{Expand YM amplitudes to YMS ones}
\label{sec-recur-constru}

In this section, we derive the color-ordered Yang-Mills (YM) amplitudes as part of the formula for expanding YM amplitudes into Yang-Mills-scalar (YMS) amplitudes. The technique employed in this section is based on the analysis of the sub-leading-order soft behavior. It represents an improvement over the method utilized in \cite{Zhou:2022orv}. In comparison with the previous method from \cite{Zhou:2022orv}, the new approach allows us to construct the expanded formula recursively, starting from the lowest $3$-point amplitudes that can be uniquely determined through bootstrapping. This is achieved without the need to rely on other frameworks, such as Feynman rules or CHY formalism. However, it should be noted that the resulting expansion in this section does not exhibit gauge invariance for all polarizations. The manifestation of gauge invariance is the objective of the next section.

\subsection{$3$-point amplitudes}
\label{subsec-3p}

To ensure our construction is self-contained, without relying on Feynman rules or the CHY formula, we fix the 3-point color-ordered Yang-Mills (YM) amplitudes using a bootstrapping method. Our construction is primarily based on the following ansatz:

1) The amplitude ${\cal A}_{\rm YM}(1,2,3)$ with the color ordering $(1,2,3)$ has a mass dimension of $1$, and due to the absence of factorization channels for the lowest-point amplitudes, it does not contain any pole structures.

2) This amplitude is linearly dependent on polarization vectors $\epsilon_1$, $\epsilon_2$ and $\epsilon_3$.

3) Due to cyclic symmetry, ${\cal A}_{\rm YM}(1,2,3)$ remains invariant under permutation transformations $(1\to2, 2\to3, 3\to1)$.

In this manner, we establish the foundational elements for our self-contained construction as

\bea
{\cal A}_{\rm YM}(1,2,3)=(k_1\cdot\epsilon_2)\,(\epsilon_3\cdot\epsilon_1)+(k_2\cdot\epsilon_3)\,(\epsilon_1\cdot\epsilon_2)
+(k_3\cdot\epsilon_1)\,(\epsilon_2\cdot\epsilon_3)\,.~~\label{3p-expli}
\eea
It's worth noting that replacing $k_1\cdot\epsilon_2$, $k_2\cdot\epsilon_3$, and $k_3\cdot\epsilon_1$ with $k_3\cdot\epsilon_2$, $k_1\cdot\epsilon_3$, and $k_2\cdot\epsilon_1$, respectively, results in an overall sign change. This occurs due to both momentum conservation and the on-shell condition $k_i\cdot\epsilon_i=0$.

We can expand ${\cal A}_{\rm YM}(1,2,3)$ onto YMS amplitudes and BAS amplitudes in the manner of equation \eqref{expan-YMS-recur}, by  ${\cal A}_{\rm YMS}(1,3;2|1,2,3)=(\epsilon_2\cdot k_1)\,{\cal A}_{\rm BAS}(1,2,3|1,2,3)$ and ${\cal A}_{\rm BAS}(1,2,3|1,2,3)=1$, we have

\bea
{\cal A}_{\rm YM}(1,2,3)&=&\big(\epsilon_3\cdot\epsilon_1\big)\,{\cal A}_{\rm YMS}(1,3;2|1,2,3)\nn
& &+\big(\epsilon_3\cdot f_2\cdot\epsilon_1\big)\,{\cal A}_{\rm BAS}(1,2,3|1,2,3)\,.~~\label{3p-expan}
\eea
Then the double copy structure indicates the expansion for general $3$-point color ordered YM amplitude
\bea
{\cal A}_{\rm YM}(\sigma_3)&=&\big(\epsilon_3\cdot\epsilon_1\big)\,{\cal A}_{\rm YMS}(1,3;2|\sigma_3)\nn
& &+\big(\epsilon_3\cdot f_2\cdot\epsilon_1\big)\,{\cal A}_{\rm BAS}(1,2,3|\sigma_3)\,,~~\label{expan-3p}
\eea
where $\sigma_3$ is an arbitrary ordering among legs in $\{1,2,3\}$. Equation \eref{expan-3p} serves as the initial step for the recursive construction in this section.

\subsection{Recursive construction for $4$-point amplitudes}
\label{subsec-4p}

In this subsection, we derive the expansion of 4-point Yang-Mills (YM) amplitudes based on the previously expanded formula for $3$-point amplitudes in equation \eqref{expan-3p}. This is accomplished by examining the sub-leading-order soft behavior of the 4-point amplitude. We label the external legs of the $4$-point amplitude ${\cal A}{\rm YM}(\sigma_4)$ as $\sigma
_4=\{1,2,3,s\}$ and choose the KK basis such that legs $1$ and $3$ are fixed at two ends in the left color ordering. We focus on the soft behavior of the external particle $s$ by rescaling $k_s$ as $k_s\to\tau k_s$ and expanding ${\cal A}{\rm YM}(\sigma_4)$ with respect to $\tau$. This process of constructing 4-point amplitudes from 3-point amplitudes provides valuable insights for addressing more general cases in our subsequent discussions.

According to the soft theorem in \eref{soft-theo-g1} and \eref{soft-fac-g-1}, the sub-leading contribution is given as
\bea
{\cal A}^{(1)_s}_{\rm YM}(\sigma_4)&=&S^{(1)_s}_g\,{\cal A}_{\rm YM}(\sigma_4\setminus s)\nn
&=&S^{(1)_s}_g\,\Big[\big(\epsilon_3\cdot\epsilon_1\big)\,{\cal A}_{\rm YMS}(1,3;2|\sigma_4\setminus s)
+\big(\epsilon_3\cdot f_2\cdot\epsilon_1\big)\,{\cal A}_{\rm BAS}(1,2,3|\sigma_4\setminus s)\Big]\nn
&=&P_1+P_2\,,~~\label{p1p2-I-5p}
\eea
The second line is derived by substituting the expansion \eref{expan-3p} into the first line. The sub-leading contribution is then split into two parts, denoted as $P_1$ and $P_2$:

\begin{align}
P_1 &= \left(\epsilon_3\cdot\epsilon_1\right) \left[S^{(1)_s}_g {\cal A}_{\rm YMS}(1,3;2|\sigma_4\setminus s)\right] + \left(\epsilon_3\cdot f_2\cdot\epsilon_1\right) \left[S^{(1)_s}_g{\cal A}_{\rm BAS}(1,2,3|\sigma_4\setminus s)\right], \nn
P_2 &= \left[S^{(1)_s}_g \left(\epsilon_3\cdot\epsilon_1\right)\right] {\cal A}_{\rm YMS}(1,3;2|\sigma_4\setminus s) + \left[S^{(1)_s}_g \left(\epsilon_3\cdot f_2\cdot\epsilon_1\right)\right] {\cal A}_{\rm BAS}(1,2,3|\sigma_4\setminus s).
\end{align}

Here, $P_1$ is obtained by applying the sub-leading soft operator $S^{(1)_s}_g$ to YMS or BAS amplitudes, and $P_2$ is obtained by acting $S^{(1)_s}_g$ on coefficients.

Our objective is to deduce the expanded formula of ${\cal A}_{\rm YM}(\sigma_4)$ from the soft behavior of ${\cal A}^{(1)s}_{\rm YM}(\sigma_4)$. Hence, it becomes imperative to construe ${\cal A}^{(1)s}_{\rm YM}(\sigma_4)$ as a synthesis of the soft behaviors inherent in the constituent components of the expansion. Here, by "components," we refer to the individual constituents akin to ${\cal A}^{(1)_s}_{\rm YMS}$ and ${\cal A}^{(0)_s}_{\rm BAS}$ within the expansion. For the $P_1$ part, the soft theorem suggests
\bea
P_1&=&\big(\epsilon_3\cdot\epsilon_1\big)\,{\cal A}^{(1)_s}_{\rm YMS}(1,3;s,2|\sigma_4)+\big(\epsilon_3\cdot f_2\cdot\epsilon_1\big)\,{\cal A}^{(1)_s}_{\rm YMS}(1,2,3;s|\sigma_4)\,.~~\label{p1-I-5p}
\eea
The second part $P_2$ can be evaluated by applying relations \eref{iden-1} and \eref{iden-2},
\bea
P_2&=&\big(\epsilon_3 \cdot f_s \cdot \epsilon_1\big)\,\Big({\delta_{1s}\over s_{1s}}+{\delta_{s3}\over s_{s3}}\Big)\,{\cal A}_{\rm YMS}(1,3;2|\sigma_4\setminus s)\nn
& &+\big(\epsilon_3\cdot f_2\cdot f_s \cdot\epsilon_1\big)\,\Big({\delta_{1s}\over s_{1s}}+{\delta_{s2}\over s_{s2}}\Big)\,{\cal A}_{\rm BAS}(1,2,3|\sigma_4\setminus s)\nn
& &+\big(\epsilon_3\cdot f_s\cdot f_2 \cdot\epsilon_1\big)\,\Big({\delta_{2s}\over s_{2s}}+{\delta_{s3}\over s_{s3}}\Big)\,{\cal A}_{\rm BAS}(1,2,3|\sigma_4\setminus s)\,.~~\label{p2-I-5p}
\eea
Since the symbol $\delta_{ab}$ will not appear in the expansion of ${\cal A}_{\rm YM}(\sigma_4)$, it should be absorbed into the soft behaviors of YMS amplitudes. By applying the soft theorem \eref{soft-theo-s} and \eref{soft-fac-s-0} for the BAS scalars, we can recognize that
\bea
& &\Big({\delta_{1s}\over s_{1s}}+{\delta_{s2}\over s_{s2}}\Big)\,{\cal A}_{\rm BAS}(1,2,3|\sigma_4\setminus s)=\tau\,{\cal A}^{(0)_s}_{\rm BAS}(1,s,2,3|\sigma_4)\,,\nn
& &\Big({\delta_{2s}\over s_{2s}}+{\delta_{s3}\over s_{s3}}\Big)\,{\cal A}_{\rm BAS}(1,2,3|\sigma_4\setminus s)=\tau\,{\cal A}^{(0)_s}_{\rm BAS}(1,2,s,3|\sigma_4)\,.~~\label{2-3-line}
\eea
This observation eliminates $\delta_{ab}$ in second and third lines in \eref{p2-I-5p}. Then we turn to the first line in \eref{p2-I-5p}.
Expanding ${\cal A}_{\rm YMS}(1,3;2|\sigma_4\setminus s)$ as in \eref{expan-YMS-recur}, we get
\bea
& &\Big({\delta_{1s}\over s_{1s}}+{\delta_{s3}\over s_{s3}}\Big)\,{\cal A}_{\rm YMS}(1,3;2|\sigma_4\setminus s)\nn
&=&(\epsilon_2\cdot k_1)\,\Big({\delta_{1s}\over s_{1s}}+{\delta_{s3}\over s_{s3}}\Big)\,{\cal A}_{\rm BAS}(1,2,3|\sigma_4\setminus s)\nn
&=&(\epsilon_2\cdot k_1)\,\Big({\delta_{1s}\over s_{1s}}+{\delta_{s2}\over s_{s2}}+{\delta_{2s}\over s_{2s}}+{\delta_{s3}\over s_{s3}}\Big)\,{\cal A}_{\rm BAS}(1,2,3|\sigma_4\setminus s)\nn
&=&\tau\,(\epsilon_2\cdot k_1)\,{\cal A}^{(0)_s}_{\rm BAS}(1,2\shuffle s,3|\sigma_4)\,.~~\label{1-line}
\eea
The second equality relies on the property $\delta_{ab}=-\delta_{ba}$, while the third equality is predicated on the soft theorem as given in equation \eref{soft-theo-s}.  Adding all of these terms together, we obtain

\bea
{\cal A}^{(1)_s}_{\rm YM}(\sigma_4)&=&\big(\epsilon_3\cdot\epsilon_1\big)\,{\cal A}^{(1)_s}_{\rm YMS}(1,3;s,2|\sigma_4)+\big(\epsilon_3\cdot f_2\cdot\epsilon_1\big)\,{\cal A}^{(1)_s}_{\rm YMS}(1,2,3;s|\sigma_4)\nn
& &+\tau\,\Big[\big(\epsilon_3\cdot f_s\cdot \epsilon_1\big)\big(\epsilon_2\cdot k_1\big){\cal A}^{(0)_s}_{\rm BAS}(1,2\shuffle s,3|\sigma_4)
+\big(\epsilon_3\cdot f_2\cdot f_s\cdot \epsilon_1\big)\,{\cal A}^{(0)_s}_{\rm BAS}(1,s,2,3|\sigma_4)\nn
& &+\big(\epsilon_3\cdot f_s\cdot f_2\cdot \epsilon_1\big)\,{\cal A}^{(0)_s}_{\rm BAS}(1,2,s,3|\sigma_4)\Big]\,.~~\label{subsoft-5p-I}
\eea

Some discussions are warranted regarding the expansion in \eref{expan-5p-I}. This expansion is derived by examining the soft behavior at the sub-leading order. The reason for selecting the sub-leading order rather than the leading one is that the leading-order contributions from the third, fourth, and fifth terms in \eref{expan-5p-I} are of order $\tau^0$, while ${\cal A}^{(0)_s}_{\rm YM}(\sigma_4)$ is of order $\tau^{-1}$. Consequently, these terms cannot be detected through the leading-order soft behavior of ${\cal A}^{(0)_s}_{\rm YM}(\sigma_4)$. This leads to a natural question: does the full expansion include a term whose leading-order soft behavior is of order $\tau^1$, making it undetectable when examining ${\cal A}^{(1)_s}_{\rm YM}(\sigma_4)$?. We can rule out this possibility through the following argument. If such a term were to exist, it must involve a coefficient that is bilinear in $k_s$, since the leading-order soft behavior of each YMS amplitude is of order $\tau^{-1}$. Consequently, the symmetry between legs $s$ and $2$ would imply the existence of another term featuring a coefficient that is bilinear in $k_2$. However, the new term with a coefficient bilinear in $k_2$ should be detectable when examining ${\cal A}^{(1)_s}_{\rm YM}(\sigma_4)$ because mass dimension considerations prohibit the coefficient from being bilinear in both $k_2$ and $k_s$. Intriguingly, the associated contribution is conspicuously absent in \eref{subsoft-5p-I}. Consequently, we can confidently assert that the hypothesized undetectable term with leading-order behavior of $\tau^1$ does not exist.

Next, we will go through each term step by step to demonstrate how to reconstruct the original amplitude ${\cal A}_{\rm YM}(\sigma_4)$ from this soft behavior.

\begin{itemize}
    \item $\big(\epsilon_3\cdot\epsilon_1\big)\,{\cal A}^{(1)_s}_{\rm YMS}(1,3;s,2|\sigma_4)$ \\
    It is evident that this term arises from the expansion represented by $\big(\epsilon_3\cdot\epsilon_1\big)\,{\cal A}_{\rm YMS}(1,3;s,2|\sigma_4)$.

    \item $\big(\epsilon_3\cdot f_2\cdot\epsilon_1\big)\,{\cal A}^{(1)_s}_{\rm YMS}(1,2,3;s|\sigma_4)$\\
    Due to the fact that $k_s$ does not contribute to the leading-order terms, the coefficients resembling $(\epsilon_3\cdot f_{2}'\cdot \epsilon_1)$ are all at leading order $\big(\epsilon_3\cdot f_2\cdot\epsilon_1\big)$. Where
    \begin{equation}
         (f'_{2})^{\mu\nu}\equiv (k_2^\mu+xk_s^\mu) \epsilon_2^\nu-\epsilon_2^\mu (k_2^\nu+y  k_s^\nu)
         \label{eq:define of f2prime}
    \end{equation}
    with $x$, $y$ can take arbitrary constant values. Therefore, this term may originate from $(\epsilon_3\cdot f_{2}'\cdot \epsilon_1){\cal A}_{\rm YMS}(1,2,3;s|\sigma_4)$ in the expansion. On the other hand, besides $\big(\epsilon_3\cdot f_2\cdot\epsilon_1\big)\,{\cal A}^{(1)_s}_{\rm YMS}(1,2,3;s|\sigma_4)$, the sub-leading order of $(\epsilon_3\cdot f_{2}'\cdot \epsilon_1){\cal A}_{\rm YMS}(1,2,3;s|\sigma_4)$ after rescale $k_s\to \tau k_s$ also contains

    \begin{equation}
        \tau\left(x(\epsilon_3\cdot k_s)(\epsilon_2\cdot \epsilon_1)-y(\epsilon_3\cdot \epsilon_2)(k_s\cdot \epsilon_1)\right){\cal A}^{(0)_s}_{\rm YMS}(1,2,3;s|\sigma_4).
    \end{equation}
    Compared with \eqref{subsoft-5p-I}, we observe that this kind of sub-leading soft behavior has not been detected, i,e, $x=y=0$. In summary, the second term arises from $\big(\epsilon_3\cdot f_2\cdot\epsilon_1\big)\,{\cal A}_{\rm YMS}(1,2,3;s|\sigma_4)$

    \item $\big(\epsilon_3\cdot f_s\cdot \epsilon_1\big)\big(\epsilon_2\cdot k_1\big){\cal A}^{(0)_s}_{\rm BAS}(1,2\shuffle s,3|\sigma_4)$\\
    The treatment of the third term, however, can be somewhat intricate. In addition, We anticipate that the expansion of ${\cal A}_{\rm YM}(\sigma_4)$ should remain invariant under the interchange of $2$ and $s$ because both $2$ and $s$ represent on-shell massless particles that have not been fixed at any ends in the color orderings. Therefore, the appearance of $\big(\epsilon_3\cdot f_2\cdot\epsilon_1\big)\,{\cal A}_{\rm YMS}(1,2,3;s|\sigma_4)$ in the expansion of ${\cal A}_{\rm YM}(\sigma_4)$ inevitably leads to the appearance of $\big(\epsilon_3\cdot f_s\cdot\epsilon_1\big)\,{\cal A}_{\rm YMS}(1,s,3;2|\sigma_4)$. On the other hand, we can derived the following relationship:

\[
{\cal A}^{(0)_s}_{\rm YMS}(1,s,3;2|\sigma_4)=(\epsilon_2\cdot k_1)\,{\cal A}^{(0)_s}_{\rm BAS}(1,2\shuffle s,3|\sigma_4).
\]

This equality can be confirmed by substituting the expansion \eref{expan-YMS-recur} and observing that $k_s$ does not contribute to $Y_2$ at the leading order. This confirms that the third term in equation \eqref{subsoft-5p-I} indeed arises from the sub-leading order soft behavior of $\big(\epsilon_3\cdot f_s\cdot\epsilon_1\big)\,{\cal A}_{\rm YMS}(1,s,3;2|\sigma_4)$.

  \item $\big(\epsilon_3\cdot f_2\cdot f_s\cdot \epsilon_1\big)\,{\cal A}^{(0)_s}_{\rm BAS}(1,s,2,3|\sigma_4)$ and $\big(\epsilon_3\cdot f_s\cdot f_2\cdot \epsilon_1\big)\,{\cal A}^{(0)_s}_{\rm BAS}(1,2,s,3|\sigma_4)$\\

  Naively, these two terms may come from $\big(\epsilon_3\cdot f'_2\cdot f_s\cdot \epsilon_1\big)\,{\cal A}^{(0)_s}_{\rm BAS}(1,s,2,3|\sigma_4)$ and $\big(\epsilon_3\cdot f_s\cdot f'_2\cdot \epsilon_1\big)\,{\cal A}^{(0)_s}_{\rm BAS}(1,2,s,3|\sigma_4)$ respectively, with $f'_2$ defined in \eqref{eq:define of f2prime}. However, from the discussion above, we realize that the coefficients in the expansion of ${\cal A}_{YM}(\sigma_4)$ do not involve terms proportional to the square of $k_s$ or higher-order terms. Therefore, these two terms exclusively originate from $\big(\epsilon_3\cdot f_2\cdot f_s\cdot \epsilon_1\big)\,{\cal A}^{(0)_s}_{\rm BAS}(1,s,2,3|\sigma_4)$ and $\big(\epsilon_3\cdot f_s\cdot f_2\cdot \epsilon_1\big)\,{\cal A}^{(0)_s}_{\rm BAS}(1,2,s,3|\sigma_4)$.

\end{itemize}

Finally, we observe the desired expansion of $4$-point amplitude

\bea
{\cal A}^{\rm I}_{\rm YM}(\sigma_4)
&=&\big(\epsilon_3\cdot\epsilon_1\big)\,{\cal A}_{\rm YMS}(1,3;s,2|\sigma_4)\nn
& &+\big(\epsilon_3\cdot f_2\cdot\epsilon_1\big)\,{\cal A}_{\rm YMS}(1,2,3;s|\sigma_4)+\big(\epsilon_3\cdot f_s\cdot \epsilon_1\big)\,{\cal A}_{\rm YMS}(1,s,3;2|\sigma_4)\nn
& &+\big(\epsilon_3\cdot f_2\cdot f_s\cdot \epsilon_1\big)\,{\cal A}_{\rm BAS}(1,s,2,3|\sigma_4)+\big(\epsilon_3\cdot f_s\cdot f_2\cdot \epsilon_1\big)\,{\cal A}_{\rm BAS}(1,2,s,3|\sigma_4)\,.~~\label{expan-5p-I}
\eea
It can be seen that this expression \eqref{expan-5p-I} remains invariant under the exchange of external legs $2$ and $s$.

\subsection{General case}
\label{subsec-general}

In this subsection, we construct the expansion of general color ordered YM amplitudes ${\cal A}_{\rm YM}(\sigma_n)$, by applying the recursive method in the previous subsection \ref{subsec-4p} iteratively.

The main result of this subsection is the expansion
\bea
{\cal A}^{\rm I}_{\rm YM}(\sigma_n)=\sum_{\vec{\pmb{\a}}}\,\big(\epsilon_n\cdot F_{\vec{\pmb{\a}}}\cdot\epsilon_1\big)\,{\cal A}_{\rm YMS}(1,\vec{\pmb{\a}},n;\{2,\cdots,n-1\}\setminus\pmb{\a}|\sigma_n)\,,~~\label{expan-form1}
\eea
where $\pmb{\a}$ denotes a subset of external legs in $\{2,\cdots,n-1\}$ which is allowed to be empty, and the ordered set $\vec{\pmb{\a}}$ is generated from $\pmb{\a}$
by endowing an order among elements in $\pmb{\a}$. The tensor $F^{\mu\nu}_{\vec{\pmb{\a}}}$ is defined as
\bea
F^{\mu\nu}_{\vec{\pmb{\a}}}\equiv\big(f_{\a_k}\cdot f_{\a_{k-1}}\cdots f_{\a_2}\cdot f_{\a_1}\big)^{\mu\nu}\,,
\eea
for $\vec{\pmb{\a}}=\{\a_1,\cdots\a_k\}$. \textbf{The summation in \eref{expan-form1} is among all un-equivalent ordered sets $\vec{\pmb{\a}}$. In other words, one should sum over all subsets
$\pmb{\a}$ of $\{2,\cdots,n-1\}$, as well as all un-cyclic permutations of elements in $\pmb{\a}$}.
Evidently, the general formula \eref{expan-form1} is satisfied by the expansions of $3$-point and $4$-point amplitudes, as shown in \eref{expan-3p} and \eref{expan-5p-I}. In the remainder of this subsection, we will illustrate that if the formula \eref{expan-form1} holds for $m$-point amplitudes, it also extends to $(m+1)$-point ones. Consequently, we can iteratively ensure the validity of the general expansion \eref{expan-form1}. This process shares many similarities with the one described in the previous subsection \ref{subsec-4p}, and as such, we will skip various details.

We can denote the external legs of the $(m+1)$-point amplitude as $s\cup{1,\cdots,m}$ and investigate the soft behavior associated with the external leg $s$. The sub-leading order soft behavior of the $(m+1)$-point amplitude is expressed as follows:
\bea
{\cal A}^{(1)_s}_{\rm YM}(\sigma_{m+1})&=&S^{(1)_s}_g\,{\cal A}^{\rm I}_{\rm YM}(\sigma_{m+1}\setminus s)\nn
&=&S^{(1)_s}_g\,\Big[\sum_{\vec{\pmb{\a}}}\,\big(\epsilon_n\cdot F_{\vec{\pmb{\a}}}\cdot\epsilon_1\big)\,{\cal A}_{\rm YMS}(1,\vec{\pmb{\a}},m;\{2,\cdots,m-1\}\setminus\pmb{\a}|\sigma_{m+1}\setminus s)\Big]\nn
&=&P_1+P_2\,,
\eea
where $P_1$ and $P_2$ are given as
\bea
P_1=\sum_{\vec{\pmb{\a}}}\,\big(\epsilon_n\cdot F_{\vec{\pmb{\a}}}\cdot\epsilon_1\big)\,\Big[S^{(1)_s}_g\,{\cal A}_{\rm YMS}(1,\vec{\pmb{\a}},m;\{2,\cdots,m-1\}\setminus\pmb{\a}|\sigma_{m+1}\setminus s)\Big]\,,
\eea
and
\bea
P_2=\sum_{\vec{\pmb{\a}}}\,\Big[S^{(1)_s}_g\,\big(\epsilon_n\cdot F_{\vec{\pmb{\a}}}\cdot\epsilon_1\big)\Big]\,{\cal A}_{\rm YMS}(1,\vec{\pmb{\a}},m;\{2,\cdots,m-1\}\setminus\pmb{\a}|\sigma_{m+1}\setminus s)\,.
\eea
The soft theorem \eref{soft-theo-g1} leads to
\bea
P_1=\sum_{\vec{\pmb{\a}}}\,\big(\epsilon_n\cdot F_{\vec{\pmb{\a}}}\cdot\epsilon_1\big)\,{\cal A}^{(1)_s}_{\rm YMS}(1,\vec{\pmb{\a}},m;s\cup\{2,\cdots,m-1\}\setminus\pmb{\a}|\sigma_{m+1})\,.~~\label{p1-I-mp}
\eea
The block $P_2$ can be calculated as
\bea
P_2&=&\sum_{\vec{\pmb{\a}}}\,\big(\epsilon_n\cdot F_{\vec{\pmb{\a}}\shuffle s}\cdot\epsilon_1\big)\,\Big({\delta_{s_ls}\over s_{s_ls}}+{\delta_{ss_r}\over s_{sr}}\Big)\,{\cal A}_{\rm YMS}(1,\vec{\pmb{\a}},m;\{2,\cdots,m-1\}\setminus\pmb{\a}|\sigma_{m+1}\setminus s)\,,~~\label{p2-step1}
\eea
by using relations in \eref{iden-1} and \eref{iden-2}. The notation $s_l$
denotes the adjacent leg of $s$ which is at the l.h.s of $s$ in $\vec{\pmb{\a}}\shuffle s$, while $s_r$ denotes the r.h.s one. Using the argument the same as that from previous subsection, we arrive at
\bea
P_2&=&\sum_{\vec{\pmb{\a}}}\,\big(\epsilon_n\cdot F_{\vec{\pmb{\a}}\shuffle s}\cdot\epsilon_1\big)\,{\cal A}^{(0)_s}_{\rm YMS}(1,\vec{\pmb{\a}}\shuffle s,m;\{2,\cdots,m-1\}\setminus\pmb{\a}|\sigma_{m+1})\,.~~\label{p2-I-mp}
\eea
Combining \eref{p1-I-mp} and \eref{p2-I-mp} together leads to
\bea
{\cal A}^{(1)_s}_{\rm YM}(\sigma_{m+1})&=&\sum_{\vec{\pmb{\a}}}\,\big(\epsilon_n\cdot F_{\vec{\pmb{\a}}}\cdot\epsilon_1\big)^{(0)_s}\,{\cal A}^{(1)_s}_{\rm YMS}(1,\vec{\pmb{\a}},m;s\cup\{2,\cdots,m-1\}\setminus\pmb{\a}|\sigma_{m+1})\nn
& &+\sum_{\vec{\pmb{\a}}}\,\big(\epsilon_n\cdot F_{\vec{\pmb{\a}}\shuffle s}\cdot\epsilon_1\big)^{(0)_s}\,{\cal A}^{(0)_s}_{\rm YMS}(1,\vec{\pmb{\a}}\shuffle s,m;\{2,\cdots,m-1\}\setminus\pmb{\a}|\sigma_{m+1})\,,
\eea
which indicates the expansion
\bea
{\cal A}_{\rm YM}(\sigma_{m+1})&=&\sum_{\vec{\pmb{\a}}}\,\big(\epsilon_n\cdot F_{\vec{\pmb{\a}}}\cdot\epsilon_1\big)\,{\cal A}_{\rm YMS}(1,\vec{\pmb{\a}},m;s\cup\{2,\cdots,m-1\}\setminus\pmb{\a}|\sigma_{m+1})\nn
& &+\sum_{\vec{\pmb{\a}}}\,\big(\epsilon_n\cdot F_{\vec{\pmb{\a}}\shuffle s}\cdot\epsilon_1\big)\,{\cal A}_{\rm YMS}(1,\vec{\pmb{\a}}\shuffle s,m;\{2,\cdots,m-1\}\setminus\pmb{\a}|\sigma_{m+1})\nn
&=&\sum_{\vec{\pmb{\a}}''}\,\big(\epsilon_n\cdot F_{\vec{\pmb{\a}}''}\cdot\epsilon_1\big)\,{\cal A}_{\rm YMS}(1,\vec{\pmb{\a}}'',m;s\cup\{2,\cdots,m-1\}\setminus\pmb{\a}''|\sigma_{m+1})\,.~~\label{expan-I-mp}
\eea
The final line in \eref{expan-I-mp} corresponds to the expanded formula \eref{expan-form1} applicable to the $(m+1)$-point case. It's worth noting that each set $\pmb{\a}$ in \eref{expan-I-mp} represents a subset of $\{2, \ldots, m-1\}$ excluding the leg $s$, while each $\pmb{\a}''$ denotes a subset of $s\cup\{2, \ldots, m-1\}$.

Hence, the general expansion formula \eref{expan-form1} is proven to be valid for $(m+1)$-point amplitudes when it has been established for $m$-point amplitudes and demonstrated for $3$-point and $4$-point amplitudes. This conclusion is derived through a mathematical induction argument.

\section{Manifests the gauge invariance}
\label{sec-GI}

The expansion presented in \eref{expan-form1} doesn't inherently exhibit gauge invariance for all polarizations, including $\epsilon_1$ and $\epsilon_n$. In this section, we establish an expansion that explicitly maintains gauge invariance through two distinct approaches. The first approach, detailed in subsection \ref{subsec-constu-dire}, involves a direct modification of \eref{expan-form1}, incorporating the gauge invariance condition. While this approach requires the imposition of gauge invariance rather than its proof, it serves as a valuable starting point. To address the logical concern associated with the first construction, we introduce the second approach in subsection \ref{subsec-recur-GI}. In this approach, we utilize the recursive technique established in the previous section \ref{sec-recur-constru}. The key difference lies in the modification of the starting point of the recursion,i.e, the $3$-point amplitude, ensuring that the expression at this initial stage exhibits explicit gauge invariance. Subsequently, at each step of the recursion, we insert soft particles while maintaining this manifest gauge invariance. Ultimately, this process allows us to construct an expansion that ensures gauge invariance without the need for explicit requirements.

\subsection{Direct construction}
\label{subsec-constu-dire}

In the expansion \eref{expan-form1}, gauge invariance is manifest for polarizations $\epsilon_i$ with $i\in\{2,\cdots,n-1\}$ because the tensor $f^{\mu\nu}_i$ automatically vanishes under the replacement $\epsilon_i\to k_i$. However, ensuring gauge invariance for polarizations $\epsilon_1$ and $\epsilon_n$ is not as straightforward. To obtain an expansion formula that also explicitly demonstrates gauge invariance for $\epsilon_1$ and $\epsilon_n$, one can impose the following gauge invariance conditions:

\bea
A_n&=&\sum_{\vec{\pmb{\a}}}\,\Big(\epsilon_n\cdot F_{\vec{\pmb{\a}}}\cdot k_1\Big)\,{\cal A}_{\rm YMS}(1,\vec{\pmb{\a}},n;\{2,\cdots,n-1\}\setminus\pmb{\a}|\sigma_n)=0\,,\nn
B_n&=&\sum_{\vec{\pmb{\a}}}\,\Big(k_n\cdot F_{\vec{\pmb{\a}}}\cdot \epsilon_1\Big)\,{\cal A}_{\rm YMS}(1,\vec{\pmb{\a}},n;\{2,\cdots,n-1\}\setminus\pmb{\a}|\sigma_n)=0\,,\nn
C_n&=&\sum_{\vec{\pmb{\a}}}\,\Big(k_n\cdot F_{\vec{\pmb{\a}}}\cdot k_1\Big)\,{\cal A}_{\rm YMS}(1,\vec{\pmb{\a}},n;\{2,\cdots,n-1\}\setminus\pmb{\a}|\sigma_n)=0\,,~~\label{gauge-inva}
\eea
which are obtained from \eref{expan-form1} by replacing $\epsilon_1$ or $\epsilon_n$ with $k_1$ or $k_n$, respectively.
These conditions lead to the following new formula,
\bea
{\cal A}^{\rm II}_{\rm YM}(\sigma_n)=-\sum_{\vec{\pmb{\a}}}\,{{\rm tr}\,\big(f_n\cdot F_{\vec{\pmb{\a}}}\cdot f_1\big)\over k_n\cdot k_1}\,{\cal A}_{\rm YMS}(1,\vec{\pmb{\a}},n;\{2,\cdots,n-1\}\setminus\pmb{\a}|\sigma_n)\,,~~\label{expan-form2}
\eea
since
\bea
{\cal A}^{\rm II}_{\rm YM}(\sigma_n)&=&{\cal A}^{\rm I}_{\rm YM}(\sigma_n)-{k_n\cdot\epsilon_1\over k_n\cdot k_1}\,A_n-{\epsilon_n\cdot k_1\over k_n\cdot k_1}\,B_n+{\epsilon_n\cdot\epsilon_1\over k_n\cdot k_1}\,C_n\,,~~\label{ABC}
\eea
where ${\cal A}^{\rm I}_{\rm YM}(\sigma_n)$ is the expanded formula in \eref{expan-form1}.

In the above construction, we required the gauge invariance for polarizations $\epsilon_1$ and $\epsilon_n$, namely, $A_n=B_n=C_n=0$. However, it is quite non-trivial to prove this property for general $n$. To ensure this gauge invariance condition, one need to derive the expansion \eref{expan-form1} from a manifestly gauge invariant framework, such as lagrangian or CHY formalism. Thus, if we insist the spirt of constructing YM amplitudes recursively from lowest-point ones, without respecting other frameworks, the above construction is not so satisfactory.

\subsection{Recursive derivation}
\label{subsec-recur-GI}

For the expansion of the simplest $3$-point amplitudes in \eref{expan-3p}, the gauge invariance for polarization $\epsilon_1$ or $\epsilon_3$ is easy to be observed. For example, replacing $\epsilon_3$ by $k_3$ in \eref{expan-form1} yields
\bea
{\cal A}^{\rm I}_{\rm YM}(\sigma_3)\Big|_{\epsilon_3\to k_3}&=&\big(k_3\cdot\epsilon_1\big)\,{\cal A}_{\rm YMS}(1,3;2|\sigma_3)\nn
& &-\big(k_3\cdot\epsilon_2\big)\,\big(k_2\cdot\epsilon_1\big)\,{\cal A}_{\rm BAS}(1,2,3|\sigma_3)\,,
\eea
here we have used $k_3\cdot k_2=0$ due to momentum conservation and on-shell condition. By utilizing \eref{expan-YMS-recur} to expand ${\cal A}_{\rm YMS}(1,3;2|\sigma_3)$ into the BAS amplitude ${\cal A}_{\rm BAS}(1,2,3|\sigma_3)$, we can readily verify that ${\cal A}^{\rm I}_{\rm YM}(\sigma_3)\Big|{\epsilon_3\to k_3}=0$. This corresponds to the gauge invariance condition stated in the second line of \eref{gauge-inva} for the case of $3$-point amplitudes. Similar verifications apply to the other two conditions in \eref{gauge-inva}. Consequently, we can transform the expansion \eref{expan-3p} into a new formulation denoted as ${\cal A}^{\rm II}_{\rm YM}(\sigma_3)$, which is expressed as:
\bea
{\cal A}^{\rm II}_{\rm YM}(\sigma_3)&=&{{\rm tr}\,\big(f_3\cdot f_1\big)\over k_3\cdot k_1}\,{\cal A}_{\rm YMS}(1,3;2|\sigma_3)\nn
& &+{{\rm tr}\,\big(f_3\cdot f_2\cdot f_1\big)\over k_3\cdot k_1}\,{\cal A}_{\rm BAS}(1,2,3|\sigma_3)\,.~~\label{expan-3p-II}
\eea

Next, we will start from expression \eqref{expan-3p-II} and repeat the procedure outlined in section \eqref{subsec-general} to iteratively construct an alternative expansion form of ${\cal A}^{\rm II}_{\rm YM}(\sigma_n)$ using lower-point amplitudes. Simultaneously, we will recursively observe that $A_n=B_n=C_n=0$ in \eqref{gauge-inva} holds for any $n$, which establishes the gauge invariance for polarizations $\epsilon_1$ and $\epsilon_n$ in the expansion \eref{expan-form1}.

First, let's make the following inductive assumption: (A) The expansion \eqref{expan-form2} that maintains manifestness of all polarization vectors is valid for all $n\leq m$. (B) The gauge invariance condition \eqref{gauge-inva} $A_n=B_n=C_n=0$ holds for all  $n\leq m$. As shown above, these two conditions hold when $n = 3$. Therefore, we have obtained the starting point for the inductive proof. We need to prove that these two conditions hold for $n = m + 1$ as well. The sub-leading soft behavior of ${\cal A}^{\rm II}_{\rm YM}(\sigma_{m+1})$ can be expressed as
\bea
{\cal A}^{(1)_s}_{\rm YM}(\sigma_{m+1})&=&S^{(1)_s}_g\,{\cal A}^{\rm II}_{\rm YM}(\sigma_{m+1}\setminus s)\nn
&=&S^{(1)_s}_g\,\Big[-\sum_{\vec{\pmb{\a}}}\,{{\rm tr}\,\big(f_m\cdot F_{\vec{\pmb{\a}}}\cdot f_1\big)\over k_m\cdot k_1}\,{\cal A}_{\rm YMS}(1,\vec{\pmb{\a}},m;\{2,\cdots,m-1\}\setminus\pmb{\a}|\sigma_{m+1}\setminus s)\Big]\nn
&=&P_1+P_2\,,~~\label{m+1p-soft-II}
\eea
where
\bea
P_1=-\sum_{\vec{\pmb{\a}}}\,{{\rm tr}\,\big(f_m\cdot F_{\vec{\pmb{\a}}}\cdot f_1\big)\over k_m\cdot k_1}\,{\cal A}^{(1)_s}_{\rm YMS}(1,\vec{\pmb{\a}},m;s\cup\{2,\cdots,m-1\}\setminus\pmb{\a}|\sigma_{m+1})
\eea
is obtained by acting the operator $S^{(1)_s}_g$ on YMS amplitudes, while $P_2$ is obtained by acting $S^{(1)_s}_g$ on coefficients.
The block $P_2$ can be evaluated as
\bea
P_2&=&-\sum_{\vec{\pmb{\a}}}\,{{\rm tr}\,\big(f_m\cdot F_{\vec{\pmb{\a}}\shuffle s}\cdot f_1\big)\over k_m\cdot k_1}\,\Big({\delta_{s_ls}\over s_{s_ls}}+{\delta_{ss_r}\over s_{ss_r}}\Big)\,{\cal A}_{\rm YMS}(1,\vec{\pmb{\a}},m;\{2,\cdots,m-1\}\setminus\pmb{\a}|\sigma_{m+1}\setminus s)\nn
& &-\sum_{\vec{\pmb{\a}}}\,{{\rm tr}\,\big(f_m\cdot F_{\vec{\pmb{\a}}}\cdot f_1\cdot f_s\big)\over k_m\cdot k_1}\,\Big({\delta_{ms}\over s_{ms}}+{\delta_{s1}\over s_{s1}}\Big)\,{\cal A}_{\rm YMS}(1,\vec{\pmb{\a}},m;\{2,\cdots,m-1\}\setminus\pmb{\a}|\sigma_{m+1}\setminus s)\nn
& &+{k_m\cdot f_s\cdot k_1\over k_m\cdot k_1}\,\Big({\delta_{1s}\over s_{1s}}+{\delta_{sm}\over s_{sm}}\Big)\,\Big[\sum_{\vec{\pmb{\a}}}\,{{\rm tr}\,\big(f_m\cdot F_{\vec{\pmb{\a}}}\cdot f_1\big)\over k_m\cdot k_1}\,{\cal A}_{\rm YMS}(1,\vec{\pmb{\a}},m;\{2,\cdots,m-1\}\setminus\pmb{\a}|\sigma_{m+1}\setminus s)\Big]\,,~~\label{P2-II}
\eea
notice that the last line arises from acting the soft operator on the denominator $k_n\cdot k_1$ in \eref{m+1p-soft-II}.
The first line in \eref{P2-II} can be recognized as
\bea
P_{21}=-\sum_{\vec{\pmb{\a}}}\,{{\rm tr}\,\big(f_m\cdot F_{\vec{\pmb{\a}}\shuffle s}\cdot f_1\big)\over k_m\cdot k_1}\,{\cal A}^{(0)_s}_{\rm YMS}(1,\vec{\pmb{\a}}\shuffle s,m;\{2,\cdots,m-1\}\setminus\pmb{\a}|\sigma_{m+1})\,,
\eea
via the technic from \eref{p2-step1} to \eref{p2-I-mp}. Then we combining the second and third lines in \eref{P2-II} and regroup them as
as
\bea
P_{22}&=&{\cal C}_1\,\Big[\sum_{\vec{\pmb{\a}}}\,\big(\epsilon_m\cdot F_{\vec{\pmb{\a}}}\cdot \epsilon_1\big)\,{\cal A}_{\rm YMS}(1,\vec{\pmb{\a}},m;\{2,\cdots,m-1\}\setminus\pmb{\a}|\sigma_{m+1}\setminus s)\Big]\nn
& &+{\cal C}_2\,\Big[\sum_{\vec{\pmb{\a}}}\,\big(\epsilon_m\cdot F_{\vec{\pmb{\a}}}\cdot k_1\big)\,{\cal A}_{\rm YMS}(1,\vec{\pmb{\a}},m;\{2,\cdots,m-1\}\setminus\pmb{\a}|\sigma_{m+1}\setminus s)\Big]\nn
& &+{\cal C}_3\,\Big[\sum_{\vec{\pmb{\a}}}\,\big(k_m\cdot F_{\vec{\pmb{\a}}}\cdot \epsilon_1\big)\,{\cal A}_{\rm YMS}(1,\vec{\pmb{\a}},m;\{2,\cdots,m-1\}\setminus\pmb{\a}|\sigma_{m+1}\setminus s)\Big]\nn
& &+{\cal C}_4\,\Big[\sum_{\vec{\pmb{\a}}}\,\big(k_m\cdot F_{\vec{\pmb{\a}}}\cdot k_1\big)\,{\cal A}_{\rm YMS}(1,\vec{\pmb{\a}},m;\{2,\cdots,m-1\}\setminus\pmb{\a}|\sigma_{m+1}\setminus s)\Big]\,,~~\label{P22-II}
\eea
where coefficients ${\cal C}_i$ with $i\in\{1,\cdots,4\}$ are independent of ordered sets $\vec{\pmb{\a}}$. Using $\delta_{ab}=\delta_{ba}$
and $k_m\cdot f_s\cdot k_1=-k_1\cdot f_s\cdot k_m$, we find ${\cal C}_1=0$. Then remaining three lines in \eref{P22-II} vanish automatically,
since the gauge invariance conditions in \eref{gauge-inva} for $m$-point amplitudes by the condition (A) of the induction hypothesis.
Thus we obtain
\bea
{\cal A}^{(1)_s}_{\rm YM}(\sigma_{m+1})&=&P_1+P_{21}\nn
&=&-\sum_{\vec{\pmb{\a}}}\,{{\rm tr}\,\big(f_m\cdot F_{\vec{\pmb{\a}}}\cdot f_1\big)\over k_m\cdot k_1}\,{\cal A}^{(1)_s}_{\rm YMS}(1,\vec{\pmb{\a}},m;s\cup\{2,\cdots,m-1\}\setminus\pmb{\a}|\sigma_{m+1})\nn
& &-\sum_{\vec{\pmb{\a}}}\,{{\rm tr}\,\big(f_m\cdot F_{\vec{\pmb{\a}}\shuffle s}\cdot f_1\big)\over k_m\cdot k_1}\,{\cal A}^{(0)_s}_{\rm YMS}(1,\vec{\pmb{\a}}\shuffle s,m;\{2,\cdots,m-1\}\setminus\pmb{\a}|\sigma_{m+1})\,,
\eea
which indicates
\bea
{\cal A}^{\rm II}_{\rm YM}(\sigma_{m+1})&=&-\sum_{\vec{\pmb{\a}}}\,{{\rm tr}\,\big(f_m\cdot F_{\vec{\pmb{\a}}}\cdot f_1\big)\over k_m\cdot k_1}\,{\cal A}_{\rm YMS}(1,\vec{\pmb{\a}},m;s\cup\{2,\cdots,m-1\}\setminus\pmb{\a}|\sigma_{m+1})\nn
& &-\sum_{\vec{\pmb{\a}}}\,{{\rm tr}\,\big(f_m\cdot F_{\vec{\pmb{\a}}\shuffle s}\cdot f_1\big)\over k_m\cdot k_1}\,{\cal A}_{\rm YMS}(1,\vec{\pmb{\a}}\shuffle s,m;\{2,\cdots,m-1\}\setminus\pmb{\a}|\sigma_{m+1})\nn
&=&-\sum_{\vec{\pmb{\a}}'}\,{{\rm tr}\,\big(f_m\cdot F_{\vec{\pmb{\a}}'}\cdot f_1\big)\over k_m\cdot k_1}\,{\cal A}_{\rm YMS}(1,\vec{\pmb{\a}}',m;s\cup\{2,\cdots,m-1\}\setminus\pmb{\a}'|\sigma_{m+1})\,.~~\label{expan-m+1p}
\eea

In the formula \eref{expan-m+1p}, $\pmb{\a}$ denotes subsets of $\{2,\cdots,m-1\}$ that do not include the leg $s$, while $\pmb{\a}'$
represents subsets of $s\cup\{2,\cdots,m-1\}$. The expansion \eref{expan-m+1p} is essentially the expanded formula ${\cal A}^{\rm II}_{\rm YM}(\sigma_n)$ as seen in \eref{expan-form2}, applied to the $(m+1)$-point case. It is important to recall that ${\cal A}^{\rm II}_{\rm YM}(\sigma_{m+1})$ and ${\cal A}^{\rm I}_{\rm YM}(\sigma_{m+1})$ are two distinct forms of expansion derived using the same method. As a result, we arrive at the conclusion that ${\cal A}^{\rm II}_{\rm YM}(\sigma_{m+1})={\cal A}^{\rm I}_{\rm YM}(\sigma_{m+1})$, leading to
\bea
-{k_m\cdot\epsilon_1\over k_m\cdot k_1}\,A_{m+1}-{\epsilon_m\cdot k_1\over k_m\cdot k_1}\,B_{m+1}+{\epsilon_m\cdot\epsilon_1\over k_m\cdot k_1}\,C_{m+1}=0\,,~~\label{gau-inv-m+1p}
\eea
where $A_{m+1}$, $B_{m+1}$, and $C_{m+1}$ are defined in \eref{gauge-inva} with $(m+1)$ replacing $n$. Consequently, we conclude that the gauge invariance conditions $A_{m+1}=B_{m+1}=C_{m+1}=0$ hold for the $(m+1)$-point amplitudes, as each coefficient associated with the independent variables
$k_m\cdot\epsilon_1$, $\epsilon_m\cdot k_1$, and $\epsilon_m\cdot\epsilon_1$ must vanish individually.

In summary, if the expansion \eref{expan-form2} and the gauge invariance conditions \eref{gauge-inva} hold for $m$-point amplitudes, they automatically hold for $(m+1)$-point amplitudes. Given that the $3$-point amplitudes ${\cal A}_{\rm YM}(\sigma_3)$ satisfy both the expansion \eref{expan-form2} and the gauge invariance conditions \eref{gauge-inva}, we can apply this result iteratively to conclude that \eref{expan-form2} and \eref{gauge-inva} are valid for any YM amplitude ${\cal A}_{\rm YM}(\sigma_n)$. In other words, we have presented an alternative expression \eref{expan-form2} that explicitly maintains gauge invariance for all polarization vectors constructed from lower-point amplitudes. Additionally, we've verified that the expansions derived in the previous section, as outlined in \eqref{subsec-general}, also preserve gauge invariance for both $\epsilon_1$ and $\epsilon_n$.

\section{Summery}
\label{sec-conclusion}

In this paper, we have enhanced the recursive method based on the sub-leading soft theorem for external gluons, as previously employed in \cite{Zhou:2022orv}. With our new approach, we have developed two types of expansions from YM amplitudes to YMS amplitudes. The first type does not exhibit manifest gauge invariance for each polarization, while the second type does. As detailed in Section \ref{sec-intro} and Section \ref{sec-background}, the sub-leading soft theorem for gluons effectively "grows" a soft gluon in a manifestly gauge-invariant manner. Consequently, as long as our recursive starting point is manifestly gauge-invariant, any intermediate result obtained during the recursion process remains explicitly gauge-invariant.

According to the methodology presented in this paper, we can also make the following two extensions:

\begin{itemize}
    \item Based on the double copy structure, one can replace YM (YMS) by GR (EYM) in \eref{expan-form1} and \eref{expan-form2} to obtain the expansion of GR amplitudes to EYM ones as follows,

\bea
{\cal A}^{\rm I}_{\rm GR}(n)=\sum_{\vec{\pmb{\a}}}\,\big(\epsilon_n\cdot F_{\vec{\pmb{\a}}}\cdot\epsilon_1\big)\,{\cal A}_{\rm EYM}(1,\vec{\pmb{\a}},n;\{2,\cdots,n-1\}\setminus\pmb{\a})\,,~~\label{expan-form3}
\eea

\bea
{\cal A}^{\rm II}_{\rm GR}(n)=-\sum_{\vec{\pmb{\a}}}\,{{\rm tr}\,\big(f_n\cdot F_{\vec{\pmb{\a}}}\cdot f_1\big)\over k_n\cdot k_1}\,{\cal A}_{\rm EYM}(1,\vec{\pmb{\a}},n;\{2,\cdots,n-1\}\setminus\pmb{\a})\,.~~\label{expan-form4}
\eea
Then, we perform a similar replacement for \eref{expan-YMS-recur-GI},

\bea
& &{\cal A}_{\rm EYM}(1,\cdots,n;\{p_1,\cdots,p_m\})\nn
&=&\sum_{\vec{\pmb{\a}}}\,{k_r\cdot F_{\vec{\pmb{\a}}}\cdot Y_{\vec{\pmb{\a}}}\over k_r\cdot k_{p_1\cdots p_m}}\,{\cal A}_{\rm EYM}(1,\{2,\cdots,n-1\}\shuffle \vec{\pmb{\a}},n;\{p_1,\cdots,p_m\}\setminus\pmb{\a})\,.~~\label{expan-YMS-recur-GI-01}
\eea
Starting from \eref{expan-form4}, iteratively using \eref{expan-YMS-recur-GI-01}, one arrives at the expansions of pure GR amplitudes to pure YM amplitudes, whose coefficients manifest the gauge invariance for each polarization. Thus the manifestly gauge invariant BCJ numerators are found.

\item  The gauge invariant expansion can be extended to the $1$-loop level straightforwardly. One can first use \eref{expan-form1} to expand the $(n+2)$-point tree YM amplitude as
\bea
{\cal A}_{\rm YM}(+,\sigma_n,-)=\sum_{\vec{\pmb{\a}}}\,(\epsilon_-\cdot F_{\vec{\pmb{a}}}\cdot\epsilon_+)\,{\cal A}_{\rm YMS}(+,\vec{\pmb{\a}},-;\{1,\cdots,n\}\setminus\pmb{\a}|+,\sigma_n,-)\,,~~\label{n+2-pt}
\eea
where two fixed legs are encoded as $+$, $-$, and $\pmb{\a}$ denotes subsets of $\{1,\cdots,n\}$. The $1$-loop amplitude can be generated by taking forward limit of \eref{n+2-pt} for legs $+$ and $-$, namely, setting $k_+=-k_-=\ell$, then gluing two legs $+$ and $-$ together by identifying $\epsilon_+$, $\epsilon_-$, and summing over all possible states \cite{He:2015yua,Cachazo:2015aol,Dong:2021qai,Zhou:2021kzv,Zhou:2022djx}. This manipulation leads to the expansion at amplitude level,
\bea
{\cal A}^{1-{\rm loop}}_{\rm YM}(\sigma_n)=\sum_{\vec{\pmb{\a}}}\, \big(\Tr\,F_{\vec{\pmb{a}}}\big)\,{\cal A}_{\rm sYMS}^{1-{\rm loop}}(\vec{\pmb{\a}};\{1,\cdots,n\}\setminus\pmb{\a}|\sigma_n)\,,~~\label{n+2-pt}
\eea
which manifests the gauge invariance. Here the subscript $s$ in ${\cal A}^{1-{\rm loop}}_{\rm sYMS}$ denotes the special type of $1$-loop YMS amplitudes those the virtual particle propagating in the loop is a scalar. At the integrand level, one can employ \eref{expan-YMS-recur-GI} to expand sYMS integrands iteratively, end with pure BAS integrands. Since $Y_{\vec{\pmb{\a}}}$ should include the loop momentum $\ell$ when applying \eref{expan-YMS-recur-GI} to the $1$-loop level, the obtained gauge invariant expansion to pure BAS ones does not hold at the amplitude level.

\end{itemize}

Moreover, in section.\ref{sec-intro}, we also pointed out that the explicit formulas of soft factors can be regarded as the consequence of the universality of soft behaviors, without respecting any top down derivation. When referring to universality, we means the soft factor for BAS scalars observed from pure BAS amplitudes holds for BAS scalars in general YMS amplitudes, and the soft factors for gluons derived from YMS amplitudes with only one external gluon also hold for general YMS amplitudes. From the traditional perspective, such universality is the consequence of the symmetries. For example, the universal soft behavior of gravitons are ensured by the asymptotic BMS symmetries of flat space time at null infinity \cite{Strominger:2013jfa,Strominger:2013lka,He:2014laa,Kapec:2014opa,Barnich1,Barnich2,Barnich3}. From the bottom up perspective, universality of soft behaviors can be taken as the basic principle, and the associated symmetries are hard to be observed. Thus, a natural question is, can we reproduce the corresponding underlying symmetries from the bottom up perspective? This is an interesting future direction.

\section*{Acknowledgments}

The authors would thank Prof. Ellis Ye Yuan, Prof. Yang Zhang and Prof. Song He for helpful suggestions.

\end{document}